# Quantum Lyapunov Control Based on the Average Value of an Imaginary Mechanical Quantity

Shuang Cong, Fangfang Meng, and Sen Kuang

*Abstract*—The convergence of closed quantum systems in the degenerate cases to the desired target state by using the quantum Lyapunov control based on the average value of an imaginary mechanical quantity is studied. On the basis of the existing methods which can only ensure the single-control Hamiltonian systems converge toward a set, we design the control laws to make the multi-control Hamiltonian systems converge to the desired target state. The convergence of the control system is proved, and the convergence to the desired target state is analyzed. How to make these conditions of convergence to the target state to be satisfied is proved or analyzed. Finally, numerical simulations for a three level system in the degenrate case transfering form an initial eigenstate to a target superposition state are studied to verify the effectiveness of the proposed control method.

## I. INTRODUCTION

In the last 30 years, the control theory of quantum systems has developed rapidly, and it was widely used in many areas. One proposed technique is the Lyapunov control [1]-[8]. There are mainly three Lyapunov functions to be selected: the Lyapunov function based on the state distance, the state error and the average value of an imaginary mechanical quantity. The so-called "imaginary mechanical quantity" means that it is a linear Hermitian operator to be designed and maybe not a physically meaningful observable such as coordinate and energy. In recent years, research results on the convergence of the control system by using the Lyapunov control method based on the average value of an imaginary mechanical quantity are as follows: The control system is asymptotically stable at the target state, if i) The internal Hamiltonian is strongly regular, i.e., the transition energies between two different levels are clearly identified; ii) The control Hamiltonians are full connected, i.e., any two levels are directly coupled [1], [4]-[8]. However, many practical systems do not satisfy these conditions which are called in the degenerate cases. For these cases, Zhao et al. utilized an implicit Lyapunov control to solve the problem of convergence for the single control Hamiltonian systems governed by the Schrödinger equation [8]. However, their proposed methods only proved that the single control Hamiltonian systems will converge toward a set, but can not ensure be asymptotically stable at the desired target state.

The aim of this paper is to make the multi-control Hamiltonian systems in the degenerate cases converge to an arbitrary target state from an arbitrary initial state. Our main contributions are as follows: i) The problem of convergence to any target eigenstate for the Schrödinger equation or any target state which commutes with the internal Hamiltonian for the quantum Liouville equation is solved by adding a restriction on the Lyapunov function and designing the implicit function perturbations. ii) The problem of convergence to the target superposition state and the target state which does not commute with the internal Hamiltonian is solved in most cases by introducing a series of constant disturbances into the control laws. iii) How to make the conditions of convergence to the target state to be satisfied are analyzed or proved.

## II. Bilinear Schrödinger Equation Case

Consider the *N*-level closed quantum system governed by the following bilinear Schrödinger equation:

$$i|\dot{\psi}(t)\rangle = (H_0 + \sum_{k=1}^{r} H_k v_k(t))|\psi(t)\rangle \qquad (1)$$

where $|\psi(t)\rangle$ is the quantum state vector, $H_0$ is the internal Hamiltonian, $H_k, (k=1,\cdots,r)$ are control Hamiltonians, and $v_k(t), (k=1,\cdots,r)$ are control laws.

Two convergence conditions for Hamiltonians in [1], [4] and [5] are i) The internal Hamiltonian is strongly regular, i.e., $\omega_{i'j'} \neq \omega_{lm}, (i',j') \neq (l,m), i',j',l,m \in \{1,2,\cdots,N\}$, where $\omega_{lm} = \lambda_l - \lambda_m$, $\lambda_l$ is the *l*-th eigenvalue of $H_0$ corresponding to the eigenstate $|\phi_l\rangle$; ii) For any $|\phi_i\rangle \neq |\phi_j\rangle$, there exists at least a *k* such that $\langle\phi_i|H_k|\phi_j\rangle \neq 0$. In order to solve the problem of convergence of the control system in the degenerate cases to the desired target state, a series of perturbations $\gamma_k(t)$, which are implicit functions of state $|\psi(t)\rangle$ and time *t*, are introduced into the control laws, then (1) becomes

$$i|\dot{\psi}(t)\rangle = (H_0 + \sum_{k=1}^{r} H_k(\gamma_k(t) + v_k(t)))|\psi(t)\rangle \qquad (2)$$

where $\gamma_k(t) + v_k(t) = u_k(t), (k=1,\cdots,r)$ are the total control laws.

Our control task is to make the control system governed by (2) transfer form an arbitrary initial pure state $|\psi_0\rangle$ to an arbitrary target pure state $|\psi_f\rangle$ by designing appropriate control laws $u_k(t) = \gamma_k(t) + v_k(t), (k=1,\cdots,r)$. In order to complete this control task, firstly, the perturbations $\gamma_k(t)$ and $v_k(t)$ are designed. Secondly, the convergence of the control system is proved. Thirdly, how to make convergence conditions to be satisfied is analyzed.

At first, let us design $\gamma_k(t), (k=1,\cdots,r)$. After introducing perturbations $\gamma_k(t)$, $H_0 + \sum_{k=1}^{r} H_k \gamma_k(t)$ can be regarded as the new internal Hamiltonian of the control system. In order to facilitate understanding the basic idea of this method, we describe the system in the eigenbasis of $H_0 + \sum_{k=1}^{r} H_k \gamma_k(t)$:

$$i|\dot{\hat{\psi}}(t)\rangle = ((\hat{H}_0 + \sum_{k=1}^{r} \hat{H}_k \gamma_k(t)) + \sum_{k=1}^{r} \hat{H}_k v_k(t))|\hat{\psi}(t)\rangle \quad (3)$$

where $|\hat{\psi}\rangle = U_1^\dagger |\psi\rangle$, $\hat{H}_0 = U_1^\dagger H_0 U_1, \hat{H}_k = U_1^\dagger H_k U_1$, $U_1 = (|\phi_{1,\gamma_1,\cdots,\gamma_r}\rangle, \cdots, |\phi_{N,\gamma_1,\cdots,\gamma_r}\rangle)$, $|\phi_{n,\gamma_1,\cdots,\gamma_r}\rangle, 1 \leq n \leq N$ are eigenstates of $H_0 + \sum_{k=1}^{r} H_k \gamma_k(t)$ corresponding to the eigenvalues $\lambda_{n,\gamma_1,\cdots,\gamma_r}$. Accordingly, $|\psi_f\rangle$ will become $|\hat{\psi}_f\rangle = U_1^\dagger |\psi_f\rangle$ which is also a functional of $\gamma_k(t)$.

The design idea of $\gamma_k(t)$ is as follows: 1) $\gamma_k(t)$ are designed to satisfy i) $\omega_{l,m,\gamma_1\cdots,\gamma_r} \neq \omega_{i,j,\gamma_1\cdots,\gamma_r}, (l,m) \neq (i,j), i,j,l,m \in \{1,2,\cdots,N\}$ holds, $\omega_{l,m,\gamma_1,\cdots,\gamma_r} = \lambda_{l,\gamma_1,\cdots,\gamma_r} - \lambda_{m,\gamma_1,\cdots,\gamma_r}$; ii) $\forall j \neq l$, for $k=1,\cdots,r$, there exists at least a $(\hat{H}_k)_{jl} \neq 0$, where $(\hat{H}_k)_{jl}$ is the (j,l)-th element of $\hat{H}_k$, thus the control system can converge toward $|\hat{\psi}_f\rangle$ by designing appropriate control laws $u_k(t) = \gamma_k(t) + v_k(t), (k=1,\cdots,r)$, thus the control system can converge toward $|\hat{\psi}_f\rangle$ by designing appropriate control laws $u_k(t) = \gamma_k(t) + v_k(t), (k=1,\cdots,r)$; 2) at the same time, $\gamma_k(t), (k=1,\cdots,r)$ themselves need converge to zero, and their convergent speed must be slower than that of the control system to $|\hat{\psi}_f\rangle$ to make $\gamma_k(t)$ take effect; 3) $\gamma_k(|\psi_f\rangle) = 0$ must hold to make the control system be asymptotically stable at the target state.

References [1], [4] and [5] proposed the restriction $V(|\psi_f\rangle) < V(|\psi_0\rangle) < V(|\psi_{other}\rangle)$ to make the system in the non-degenerate cases converge to the target state $|\psi_f\rangle$ from the initial state $|\psi_0\rangle$, where $|\psi_{other}\rangle$ represents any other state in the invariant set in

$E = \{|\psi\rangle | \dot{V}(|\psi\rangle) = 0\}$ except the target state. However, in fact it is difficult to design the imaginary mechanical quantity to make this restriction on the Lyapunov function be satisfied for any initial state and any target state.

For the degenerate cases, in order to make the the system converge to the target state, we choose a simpler restriction: $V(|\psi_f\rangle) < V(|\psi_{other}\rangle)$ which can be satisfied for any initial state and any target state by designing the imaginary mechanical quantity. In order to ensure the the system converge to the target state by adding this retriction, we design all the perturbations $\gamma_k(t) = 0$ holds for $k = 1, \cdots, r$ only at $|\psi_f\rangle$, i.e., 1) $\gamma_k(|\psi_f\rangle) = 0, (k = 1, \cdots, r)$, and 2) for $|\psi\rangle \neq |\psi_f\rangle$, there exists at least one $k$ such that $\gamma_k(|\psi\rangle) \neq 0$.

According to the analysis mentioned above, let us design the specific $\gamma_k(t), (k = 1, \cdots, r)$. Since the evolution of the system's state relies on the continuous decrease of the Lyapunov function $V(t)$ in the Lyapunov control, we design $\gamma_k(t)$ be a monotonically increasing functional of $V(t)$ as:

$$\gamma_k(|\psi\rangle) = C_k \cdot \theta_k(V(|\psi\rangle) - V(|\psi_f\rangle)) \tag{4}$$

where $C_k \geq 0$, and for $k = 1, \cdots, r$, there exists at least a $C_k > 0$. And $\theta_k(\cdot)$ satisfies $\theta_k(0) = 0$, $\theta_k(s) > 0$ and $\theta_k'(s) > 0$ for every $s > 0$. In this note, the specific Lyapunov function based on the average value of an imaginary mechanical quantity is selected as：

$$V(|\psi\rangle) = \langle \psi | P_{\gamma_1, \cdots, \gamma_r} | \psi \rangle \tag{5}$$

where $P_{\gamma_1, \cdots, \gamma_r} = f(\gamma_1(t), \cdots, \gamma_r(t))$ is a functional of $\gamma_k(t)$ and positive definite.

The existence of $\gamma_k(t)$ can be established by Lemma 1.

**Lemma 1:** If $C_k = 0$, $\gamma_k(|\psi\rangle) = 0$. Else if $C_k > 0$, $\theta_k \in C^\infty(R^+; [0, \gamma_k^*]), k = 1, \cdots, r$ ($\gamma_k^*$ is a positive constant) satisfy $\theta_k(0) = 0$, $\theta_k(s) > 0$ and $\theta_k'(s) > 0$ for every $s > 0$, and $|\theta_k'| < 1/(2C_k C^*)$, $C^* = 1 + C$, $C = \max\{\|\partial P_{\gamma_1, \cdots, \gamma_r} / \partial \gamma_k\|_\infty, \gamma_k \in [0, \gamma_k^*]\}$, then for every $|\psi\rangle \in S^{2N-1}$, there is a unique $\gamma_k \in C^\infty(\gamma_k \in [0, \gamma_k^*])$ satisfying

$$\gamma_k(|\psi\rangle) = C_k \cdot \theta_k(\langle\psi|P_{\gamma_1,\cdots,\gamma_r}|\psi\rangle - \langle\psi_f|P_{\gamma_1,\cdots,\gamma_r}|\psi_f\rangle) \qquad (6)$$

**Proof:**

Assume $P_{\gamma_1,\cdots,\gamma_r}$ are analytic functions of the parameters $\gamma_k(\psi) \in [0, \gamma_k^*], (k=1,\cdots,r)$. $\partial P_{\gamma_1,\cdots,\gamma_r}/\partial \gamma_k$ are bounded on $[0, \gamma_k^*]$, thus $C < \infty$. Define

$$F_k(\gamma_1,\cdots,\gamma_r,|\psi\rangle) = \gamma_k - C_k \cdot \theta_k(\langle\psi|P_{\gamma_1,\cdots,\gamma_r}|\psi\rangle - \langle\psi_f|P_{\gamma_1,\cdots,\gamma_r}|\psi_f\rangle)$$

where $F_k(\gamma_1,\cdots,\gamma_r,|\psi\rangle)$ are regular. For a fixed $|\psi\rangle$, $F_k(\gamma_1(|\psi\rangle),\cdots,\gamma_r(|\psi\rangle),|\psi\rangle) = 0$ holds. Some deductions show that $\partial F_k(\gamma_1,\cdots,\gamma_r,|\psi\rangle)/\partial \gamma_k \neq 0$ holds. Thus according to the implicit function Theorem [8], Lemma 1 is proved. □

**Remark 1:** For the sake of simplicity, set $\gamma_k(t)=0$ for some $k$, and other $\gamma_k(t)$ are equal, denoted by $\gamma(t)$, i.e., set

$$\begin{aligned}&\gamma_k(t)=\gamma(t)=\theta(\langle\psi|P_\gamma|\psi\rangle - \langle\psi_f|P_\gamma|\psi_f\rangle), k=k_1,\cdots,k_m; \\ &\gamma_k(t)=0, k\neq k_1,\cdots,k_m (1\leq k_1,\cdots,k_m \leq r)\end{aligned} \qquad (7)$$

where $\theta(\cdot) = \theta_{k_1}(\cdot) = \cdots = \theta_{k_m}(\cdot)$ and $P_\gamma$ are functionals of $\gamma(t)$.

Then let us design $v_k(t)$ to make $\dot{V}(t) \leq 0$ holds. Setting $[P_\gamma, H_0 + \sum_{n=k_1}^{k_m} H_n \gamma(t)] = 0$, one can obtain the time derivative of the selected Lyapunov function as:

$$\begin{aligned}\dot{V} = &\sum_{k=1}^{r} iv_k(t)\langle\psi|[H_k, P_\gamma]|\psi\rangle \cdot (1+\theta'(\langle\psi_f|(\partial P_\gamma/\partial\gamma)|\psi_f\rangle)) \\ &/(1-\theta'(\langle\psi|(\partial P_\gamma/\partial\gamma)|\psi\rangle - \langle\psi_f|(\partial P_\gamma/\partial\gamma)|\psi_f\rangle))\end{aligned} \qquad (8)$$

According to Lemma 1, one can obtain $(1+\theta'(\langle\psi_f|(\partial P_\gamma/\partial\gamma)|\psi_f\rangle))/(1-\theta'(\langle\psi|(\partial P_\gamma/\partial\gamma)|\psi\rangle - \langle\psi_f|(\partial P_\gamma/\partial\gamma)|\psi_f\rangle)) > 0$ holds. In order to ensure $\dot{V}(t) \leq 0$, $v_k(t), (k=1,\cdots,r)$ are designed as:

$$v_k(t) = -K_k f_k\left(i\langle\psi|[H_k, P_\gamma]|\psi\rangle\right), (k=1,\cdots,r) \qquad (9)$$

where $K_k$ is a constant and $K_k > 0$, and $y_k = f_k(x_k), (k=1,2,\cdots,r)$ are monotonic increasing functions through the coordinate origin of the plane $x_k - y_k$.

Based on LaSalle's invariance principle [9], the convergence of the control system governed by (2) can be obtained as follows:

**Theorem 1**: Consider the control system governed by (2) with control fields $u_k(t) = \gamma_k(t) + v_k(t), (k=1,\cdots,r)$, where $\gamma_k(t)$ defined by Lemma 1 and (7), and $v_k(t)$ defined by (9). If the control system satisfies: i) $\omega_{l,m,\gamma} \neq \omega_{i,j,\gamma}, (l,m) \neq (i,j)$, $i,j,l,m \in \{1,2,\cdots,N\}$, $\omega_{l,m,\gamma} = \lambda_{l,\gamma} - \lambda_{m,\gamma}$, where $\lambda_{l,\gamma}$ is the $l$-th eigenvalue of $H_0 + \sum_{n=k_1}^{k_m} H_n \gamma(t)$ corresponding to the eigenstate $|\phi_{l,\gamma}\rangle$; ii) For any $i \neq j$, $i,j \in \{1,2,\cdots,N\}$, there exits at least one $k$ such that $(\hat{H}_k)_{lm} \neq 0$, where $(\hat{H}_k)_{lm}$ is the $(l,m)$-th element of $\hat{H}_k = U_1^\dagger H_k U_1$, $U_1 = (|\phi_{1,\gamma}\rangle, \cdots, |\phi_{N,\gamma}\rangle)$; iii) $[P_\gamma, H_0 + \sum_{n=k_1}^{k_m} H_n\gamma(t)] = 0$; iv) $(\hat{P}_\gamma)_{ll} \neq (\hat{P}_\gamma)_{mm}, l \neq m$, where $(\hat{P}_\gamma)_{ll}$ is the $(l,l)$-th element of $U_1^\dagger P_\gamma U_1$, then any trajectory will converge toward $E_1 = \left\{ |\psi_{t_0}\rangle \Big| e^{i\theta_l} \big|\phi_{l,\gamma(|\psi_{t_0}\rangle)}\big\rangle; \theta_l \in R, l \in \{1,\cdots,N\}\right\}$.

**Proof:**

Without loss of generality, assume that for $t \geq t_0, (t_0 \in R)$, $\dot{V} = 0$ is satisfied. By (8) and (9), one obtains

$$\dot{V} = 0 \Leftrightarrow \langle \psi |[H_k, P_\gamma]|\psi\rangle = 0 \Leftrightarrow v_k(t) = 0 \tag{10}$$

As $\dot{V} = 0$, $\gamma$ is a constant, denoted by $\bar{\gamma}$. The state $|\psi(t_0)\rangle$ can be written as $|\psi(t_0)\rangle = \sum_{l=1}^N c_l(t_0)|\phi_{l,\gamma}\rangle$. Then $|\hat{\psi}(t_0)\rangle$ can be written as $|\hat{\psi}(t_0)\rangle = \sum_{l=1}^N c_l(t_0) U_1^H |\phi_{l,\gamma}\rangle$. Substituting the solution of (3) with $\gamma = \bar{\gamma}$ and $v_k(t) = 0$ into $\langle \hat{\psi}|[\hat{H}_k, \hat{P}_\gamma]|\hat{\psi}\rangle = \langle \psi|[H_k, P_\gamma]|\psi\rangle = 0$, gives

$$\sum_{l,m=1}^N e^{i\omega_{l,m,\bar{\gamma}}(t-t_0)} \left((\hat{P}_{\bar{\gamma}})_{mm} - (\hat{P}_{\bar{\gamma}})_{ll}\right) c_l^*(t_0) c_m(t_0) (\hat{H}_k)_{lm} = 0 \tag{11}$$

By conditions i)-ii) and iv), one can have

$$c_l^*(t_0) c_m(t_0) = 0, (l, m \in \{1,\cdots,N\}) \tag{12}$$

which implies that there is at most one $c_l(t_0)(l \in \{1,\cdots,N\})$ which is nonzero. Theorem 1 is proved.□

Theorem 1 guarantees the control system converges to the set $E_1$, however, it can not guarantee the control system converges to the target state. From Theorem 1, we can see that if the target state $|\psi_f\rangle$ is an eigenstate, $|\psi_f\rangle$ is contained in $E_1$ because of $\gamma(|\psi_f\rangle) = 0$. In order to make the system converge to the target state, on the one hand, as $\dot{V} \leq 0$, we design $P_\gamma$ to make

$$V(|\psi_f\rangle) < V(|\psi_{other}\rangle) \tag{13}$$

hold, where $|\psi_{other}\rangle$ represents any other state in the set $E_1$ except the target state. On the other hand, because $\partial \gamma / \partial V > 0, \dot{V} \leq 0, \gamma \geq 0$ holds, we set $\gamma = \bar{\gamma} - \alpha, (0 < \alpha << \bar{\gamma})$ when $v_k(t) = 0, \gamma(t) = \bar{\gamma} \neq 0$ holds for some time to make the state trajectory evolve but not stay in $E_1$ until $|\psi_f\rangle e^{i\theta_l}$, which is the equivalent state of target state $|\psi_f\rangle$, is reached.

From the above analysis, we can see that if the control system satisfies the conditions i)-iv) in Theorem 1 and Eq.(13), and at the same time set $\gamma = \bar{\gamma} - \alpha, (0 < \alpha << \bar{\gamma})$ when $v_k(t) = 0, \gamma(t) = \bar{\gamma} \neq 0$ holds for some time, the control system can converge to the target eigenstate from an arbitrary initial pure state.

Next we'll analyze how to make these conditions be satisfied in detail. Conditions i) and ii) in Theorem 1 are associated with $H_0$, $H_k, (k = 1, \cdots, r)$ and $\gamma_k(t)$. By designing appropriate $\gamma_k(t)$, these two conditions can be satisfied in most cases. Condition iii) means that $P_\gamma$ and $H_0 + \sum_{n=k_1}^{k_m} H_n \gamma(t)$ have the same eigenstates. We design the eigenvalues of $P_\gamma$ be constant, denoted by $P_1, P_2, \cdots, P_N$, and design $P_\gamma$ as

$$P_\gamma = \sum_{j=1}^{N} P_j |\phi_{j,\gamma}\rangle\langle\phi_{j,\gamma}| \tag{14}$$

then condition iii) can be satisfied. If design $P_l \neq P_j (\forall l \neq j; 1 \leq l, j \leq N)$ to make condition iv) hold. Then let us analyze how to make (13) hold. The research result is given by the following Theorem 2.

**Theorem 2:** If one designs $P_i > P_f, (i = 1, \cdots, N, P_i \neq P_f)$, then $V(|\psi_f\rangle) < V(|\psi_{other}\rangle)$ holds, where $P_f$ is the eigenvalue of $P_{\gamma(|\psi_f\rangle)}$ corresponding to $|\psi_f\rangle$.

**Proof**:

Set $|\psi_s\rangle = \left(e^{i\theta_l}|\phi_{l,\gamma}\rangle\right)\Big|_{\gamma=0}$. According to Proposition 1 in [8], if one designs $P_i > P_f, (i = 1, \cdots, N, P_i \neq P_f)$, then $V(|\psi_f\rangle) < V(|\psi_s\rangle)$ holds. Because of

$\partial \gamma / \partial V > 0, \dot{V} \leq 0, \gamma > 0$, $V(|\psi_s\rangle) < V(|\psi_{other}\rangle)$ holds. Thus $V(|\psi_f\rangle) < V(|\psi_{other}\rangle)$ holds. Thereom 2 is proved. □

**Remark 2:** According to the above analysis and Theorem 2, the design principle of the imaginary mechanical quantity is $P_i > P_f, (i = 1, \cdots, N, P_i \neq P_f)$ and $P_l \neq P_j (\forall l \neq j)$.

In order to solve the problem of convergence to the target state being a superposition state, a series of disturbances $\eta_k$ whose values are constant are introduced into the control laws. Thus the mechanical equation (2) will become

$$i|\dot{\psi}(t)\rangle = (H_0 + \sum_{k=1}^{r} H_k(\eta_k + \gamma_k(t) + v_k(t)))|\psi(t)\rangle \quad (15)$$

Our basic idea is to design $\eta_k$ to make the target state $|\psi_f\rangle$ be an eigenstate of $H_0' = H_0 + \sum_{k=1}^{r} H_k \eta_k$. $H_0'$ can be viewed as the new internal Hamiltonian of the control system. If the number of the control Haimltonians $r$ is large enough, by designing appropriate $\eta_k$, $(H_0 + \sum_{k=1}^{r} H_k \eta_k)|\psi_f\rangle = \lambda_f' |\psi_f\rangle$ can be satisfied in most cases, where $\lambda_f'$ is the eigenvalue of $H_0'$ corresponding to $|\psi_f\rangle$. Then the design of control laws and the convergence proof can follow the target eigenstate cases. One can prove that the designed control laws are also valid and Theorem 1 and Theorem 2 also holds with changing $H_0$ into $H_0'$.

### III. QUANTUM LIOUVILLE EQUATION CASE

Consider the *N*-level closed quantum system governed by the following quantum Liouville equation:

$$i\dot{\rho}(t) = [H_0 + \sum_{k=1}^{r} H_k(\gamma_k(t) + v_k(t) + \eta_k), \rho(t)] \quad (16)$$

where $\gamma_k(t) + v_k(t) + \eta_k = u_k(t), (k = 1, \cdots, r)$ are the total control laws.

The design ideas are similar to that of Section II. The specific Lyapunov function is selected as:

$$V(\rho) = tr(P_{\gamma_1, \cdots, \gamma_r} \rho) \quad (17)$$

where $P_{\gamma_1, \cdots, \gamma_r} = f(\gamma_1(t), \cdots, \gamma_r(t))$ is a functional of $\gamma_k(t)$ and positive definite.

For the sake of simplicity, design $\gamma_k(t)$ as

$$\begin{aligned} &\gamma_k(t) = \gamma(t) = \theta(V(\rho) - V(\rho_f)), k = k_1, \cdots, k_m; \\ &\gamma_k(t) = 0, k \neq k_1, \cdots, k_m (1 \leq k_1, \cdots, k_m \leq r) \end{aligned} \quad (18)$$

where $\theta(\cdot)$ satisfies $\theta(0)=0$, $\theta(s)>0$ and $\theta'(s)>0$ for every $s>0$. Accordingly, $P_{\gamma_1,\cdots,\gamma_r}$ becomes $P_\gamma$. The existence of $\gamma(t)$ can be depicted by Lemma 2.

**Lemma 2:** If $\theta \in C^\infty(R^+;[0,\gamma^*]), k=1,\cdots,r$ ($\gamma^*$ is a positive constant) satisfy $\theta(0)=0$, $\theta(s)>0$ and $\theta'(s)>0$ for every $s>0$, and $|\theta'|<1/(2C^*)$, $C^*=1+C$, $C=\max\{\|\partial P_\gamma/\partial\gamma\|_{m_1}, \gamma \in [0,\gamma^*]\}$, then for every $\rho$, there is a unique $\gamma \in C^\infty (\gamma \in [0,\gamma^*])$ satisfying $\gamma(\rho)=\theta(tr(P_\gamma\rho)-tr(P_\gamma\rho_f))$.

The idea of proof is similar to that of Lemma 1 in Section II.

Then let us design $v_k(t)$ such that $\dot{V}(t) \le 0$ holds. Setting $[P_\gamma, H_0+\sum_{k=1}^{r}H_k\eta_k+\sum_{n=k_1}^{k_m}H_n\gamma(t)]=0$, one can obtain

$$\dot{V} = -(1+\theta'tr((\partial P_\gamma/\partial\gamma)\rho_f))/(1-\theta'tr((\partial P_\gamma/\partial\gamma)(\rho-\rho_f)))\cdot \sum_{k=1}^{r}itr([P_\gamma,H_k]\rho)v_k(t) \qquad (19)$$

By $|\theta'|<1/(2C^*)$ in Lemma 2, $(1+\theta'tr((\partial P_\gamma/\partial\gamma)\rho_f))/(1-\theta'tr((\partial P_\gamma/\partial\gamma)(\rho-\rho_f)))>0$ holds. In order to ensure $\dot{V}(t)\le 0$, $v_k(t), (k=1,\cdots,r)$ are designed as:

$$v_k(t) = K_k f_k\left(itr([P_\gamma,H_k]\rho)\right), (k=1,\cdots,r) \qquad (20)$$

where $K_k$ is a constant and $K_k>0$, and $y_k=f_k(x_k), (k=1,2,\cdots,r)$ are monotonic increasing functions which are through the coordinate origin of the plane $x_k-y_k$.

Based on LaSalle's invariance principle, the convergence of the control system can be obtained as follows.

**Theorem 3:** Consider the control system depicted by (16) with control laws $\gamma_k(t)$ defined by Lemma 2 and Eq. (18), and $v_k(t)$ defined by (20). If the control system satisfies: i) $\omega_{l,m,\gamma} \ne \omega_{i,j,\gamma}, (l,m)\ne(i,j), i,j,l,m \in \{1,2,\cdots,N\}$, $\omega_{l,m,\gamma}=\lambda_{l,\gamma}-\lambda_{m,\gamma}$, where $\lambda_{l,\gamma}$ is the $l$-th eigenvalue of $H_0+\sum_{k=1}^{r}H_k\eta_k+\sum_{n=k_1}^{k_m}H_n\gamma(t)$ corresponding to the eigenstate $|\phi_{l,\gamma}\rangle$; ii) $\forall j \ne l$, for $k=1,\cdots,r$, there exists at least a $(\hat{H}_k)_{jl} \ne 0$, where $(\hat{H}_k)_{jl}$ is the $(j,l)$-th element of $\hat{H}_k=U_2^\dagger H_k U_2$ with $U_2=(|\phi_{1,\gamma}\rangle,\cdots,|\phi_{N,\gamma}\rangle)$; iii)

$[P_\gamma, H_0+\sum_{k=1}^{r}H_k\eta_k+\sum_{n=k_1}^{k_m}H_n\gamma(t)]=0, 1\le k_1,\cdots,k_m\le r$ ; iv) For any $l\ne j, (1\le l,j\le N)$, $(\hat{P}_\gamma)_{ll}\ne(\hat{P}_\gamma)_{jj}$ holds, where $(\hat{P}_\gamma)_{ll}$ is the $(l,l)$-th element of $\hat{P}_\gamma=U_2^\dagger P_\gamma U_2$, then the control system will converge toward $E_2=\left\{\rho_{t_0}\left|\left(U_2^\dagger \rho_{t_0} U_2\right)_{ij}=0, \gamma=\gamma(\rho_{t_0}), t_0\in R\right.\right\}$.

**Proof**:

Without loss of generality, assume that for $t\ge t_0, (t_0\in R)$, $\dot{V}=0$ is satisfied. By (19) and (20), one can get

$$\dot{V}=0 \Leftrightarrow tr([P_\gamma, H_k]\rho)=0 \Leftrightarrow v_k(t)=0 \tag{21}$$

As $\dot{V}=0$, $\gamma$ are constants, denoted by $\bar{\gamma}$. The control system in the eigenbasis of $H_0+\sum_{k=1}^{r}H_k\eta_k+\sum_{n=k_1}^{k_m}H_n\gamma(t)$ is

$$i\dot{\hat{\rho}}(t)=[(\hat{H}_0+\sum_{k=1}^{r}\hat{H}_k(\gamma_k(t)+\eta_k))+\sum_{k=1}^{r}\hat{H}_k v_k(t), \hat{\rho}(t)] \tag{22}$$

where $\hat{\rho}=U_2^\dagger \rho U_2, \hat{H}_0=U_2^\dagger H_0 U_2, \hat{H}_k=U_2^\dagger H_k U_2$. Set $\hat{\rho}_{t_0}=\hat{\rho}(t_0)$. Substituting the solution of Eq. (22) with $\gamma_k(t)$ defined by Eq. (18), $\gamma=\bar{\gamma}$, and $v_k(t)=0$ into $tr([\hat{P}_{\bar{\gamma}}, \hat{H}_k]\hat{\rho})=tr([P_\gamma, H_k]\rho)=0$, gives

$$tr(e^{-i(\hat{H}_0+\sum_{k=1}^{r}H_k\eta_k+\sum_{n=k_1}^{k_m}\hat{H}_n\bar{\gamma})(t-t_0)}\hat{\rho}_{t_0}e^{i(\hat{H}_0+\sum_{k=1}^{r}H_k\eta_k+\sum_{n=k_1}^{k_m}\hat{H}_n\bar{\gamma})(t-t_0)}[\hat{P}_{\bar{\gamma}}, \hat{H}_k])=0 \tag{23}$$

where $\hat{P}_{\bar{\gamma}}=U_2^\dagger P_{\bar{\gamma}} U_2$. By condition iii), one can obtain

$$\sum_{j,l=1}^{N}\omega_{j,l,\bar{\gamma}}^n(\hat{H}_k)_{jl}((\hat{P}_{\bar{\gamma}})_{ll}-(\hat{P}_{\bar{\gamma}})_{jj})(\hat{\rho}_{t_0})_{lj}=0 \tag{24}$$

Set

$$\xi_k=\begin{bmatrix}(\hat{H}_k)_{12}((\hat{P}_{\bar{\gamma}})_{22}-(\hat{P}_{\bar{\gamma}})_{11})(\hat{\rho}_{t_0})_{21}\\ \vdots \\ (\hat{H}_k)_{(N-1)N}((\hat{P}_{\bar{\gamma}})_{NN}-(\hat{P}_{\bar{\gamma}})_{(N-1)(N-1)})(\hat{\rho}_{t_0})_{N(N-1)}\end{bmatrix},$$
$$\Lambda=diag(\omega_{1,2,\bar{\gamma}},\cdots,\omega_{N-1,N,\bar{\gamma}}), \tag{25}$$
$$M=\begin{bmatrix}1 & 1 & \cdots & 1\\ \omega_{1,2,\bar{\gamma}}^2 & \omega_{1,3,\bar{\gamma}}^2 & \cdots & \omega_{N,N-1,\bar{\gamma}}^2\\ \vdots & \vdots & \vdots & \vdots \\ \omega_{1,2,\bar{\gamma}}^{N(N-1)-2} & \omega_{1,3,\bar{\gamma}}^{N(N-1)-2} & \cdots & \omega_{N,N-1,\bar{\gamma}}^{N(N-1)-2}\end{bmatrix}$$

For $n=0,2,4,\cdots$, (24) reads $M\Im(\xi_k)=0$. For $n=1,3,5,\cdots$, (24) reads $M\Lambda\Re(\xi_k)=0$. By

condition i), and $M$ and $\Lambda$ are nonsingular real matrices, one can obtain $\xi_k = 0$. By condition ii) and iv), one have $(\hat{\rho}_{t_0})_{lj} = 0$ holds. Theorem 3 is proved. □

If $r$ is large enough, by designing appropriate $\eta_k$, $\left[H_0 + \sum_{k=1}^{r} H_k \eta_k, \rho_f\right] = 0$ can be satisfied in most cases. Then the target state $\rho_f$ is contained in $E_2$. For the special case that the target state commutes with the internal Hamiltonian, i.e., $[\rho_f, H_0] = 0$, set $\eta_k = 0$. Some analyses show that $E_2$ has at most $N!$ elements. In order to make the system converge to the target state $\rho_f$, on the one hand, we design $P_\gamma$ to make

$$V(\rho_f) < V(\rho_{other}) \qquad (26)$$

hold, where $\rho_{other}$ represents any other state in the set $E_2$ except the target state. On the other hand, we design $\gamma = \bar{\gamma} - \alpha, (0 < \alpha \ll \bar{\gamma})$ when $v_k(t) = 0, \gamma(t) = \bar{\gamma} \neq 0$ holds for some time to make the state trajectory evolve but not stay in $E_2$ until $\rho_f$ is reached.

Next we'll analyze how to make these conditions be satisfied. For satisfaction of conditions i) - iv), one can follow that of Section II. Then let us analyze how to make (26) hold. Denoting the eigenstates of $H_0 + \sum_{k=1}^{r} H_k \eta_k$ as $|\phi_{i,\eta}\rangle, (i \in \{1, \cdots, N\})$, $\tilde{\rho}_f = U_3^\dagger \rho_f U_3$ can be expressed by a diagonal matrix, where $U_3 = (|\phi_{1,\eta}\rangle, \cdots, |\phi_{N,\eta}\rangle)$. The research result is as follows:

**Theorem 4:** If $(\tilde{\rho}_f)_{ii} < (\tilde{\rho}_f)_{jj}, 1 \leq i, j \leq N$, design $P_i > P_j$; if $(\tilde{\rho}_f)_{ii} = (\tilde{\rho}_f)_{jj}, 1 \leq i, j \leq N$, design $P_i \neq P_j$; else if $(\tilde{\rho}_f)_{ii} > (\tilde{\rho}_f)_{jj}, 1 \leq i, j \leq N$, design $P_i < P_j$, then $V(\rho_f) < V(\rho_{other})$ holds, where $(\tilde{\rho}_f)_{ii}$ is the $(i,i)$-th element of $\tilde{\rho}_f$.

**Proof:**

At first, propositions 1 and 2 are proposed, then Theorem 4 are proved according to these two propositions.

**Proposition 1:** If $\{(\tilde{\rho}_f)_{11}, (\tilde{\rho}_f)_{22}, \cdots, (\tilde{\rho}_f)_{NN}\}$ arranged in a decreasing order, design $\{P_1, P_2, \cdots, P_N\}$ arranged in an increasing order, then $V(\rho_f) < V(\rho_{other})$ holds.

**Proof:**

Denote $\tilde{\rho}_s = U_3^\dagger \rho_s U_3 = diag((\tilde{\rho}_f)_{11(\tau)}, (\tilde{\rho}_f)_{22(\tau)}, \cdots, (\tilde{\rho}_f)_{NN(\tau)})$, where $\{11(\tau), 22(\tau), \cdots, NN(\tau)\}$ is a permutation of $\{11, 22, \cdots, NN\}$. At first, we prove $V(\rho_f) < V(\rho_s)$. The Lyapunov function $V(\rho) = tr(P_\gamma \rho)$ for $\gamma = 0$ can be written as

$$V(\rho)|_{\gamma=0} = \sum_{j=1}^{N} P_j \tilde{\rho}_{jj} \tag{27}$$

where $\tilde{\rho}_{jj}$ is the (j,j)-th element of $\tilde{\rho} = U_3^\dagger \rho U_3$. Assume $(\tilde{\rho}_f)_{11} \cdots > (\tilde{\rho}_f)_{NN} \geq 0$, and $0 < P_1 < \cdots < P_N$.

For N=2,

$$V(\rho_f)_2 - V(\rho_s)_2 = (P_1 - P_2)((\tilde{\rho}_f)_{11} - (\tilde{\rho}_f)_{22}) < 0 \tag{28}$$

where the subscript "2" in $V(\rho_f)_2$ and $V(\rho_s)_2$ means $N = 2$. Proposition 1 is true.

Assume Proposition 1 is true for N-1. Then

$$V(\rho_f)_{N-1} - V(\rho_s)_{N-1} = \sum_{j=1}^{N-1} P_j((\tilde{\rho}_f)_{jj} - (\tilde{\rho}_f)_{jj(\tau)}) = \sum_{j=1}^{N-1}(P_{j(\tau)} - P_j)(\tilde{\rho}_f)_{jj(\tau)} < 0 \tag{29}$$

Where $P_{j(\tau)} = (P_\gamma|_{\gamma=0})_{jj(\tau)}$.

For N,

$$V(\rho_f)_N - V(\rho_s)_N = \sum_{j=1}^{N-1}(P_{j(\tau)} - P_j)(\tilde{\rho}_f)_{jj(\tau)} + (P_{N(\tau)} - P_N)(\tilde{\rho}_f)_{NN(\tau)} \tag{30}$$

By (29) and $0 < P_1 < P_2 < \cdots < P_N$, one can get

$$V(\rho_f)_N - V(\rho_s)_N < 0 \tag{31}$$

Because of $\partial \gamma / \partial V > 0, \dot{V} \leq 0, \gamma > 0$, $V(\rho_s) < V(\rho_{other})$ holds. Thus Proposition 1 is proved. □

**Proposition 2:** If the diagonal elements of the diagonal target state $\{(\rho_f)_{11}, (\rho_f)_{22}, \cdots, (\rho_f)_{NN}\}$ are arranged in a non-decreasing order with

$(\rho_f)_{k_{11}k_{11}} = \cdots = (\rho_f)_{k_{1L_1}k_{1L_1}} < (\rho_f)_{k_{21}k_{21}} = \cdots = (\rho_f)_{k_{2L_2}k_{2L_2}}$
$< \cdots < (\rho_f)_{k_{Q1}k_{Q1}} = \cdots = (\rho_f)_{k_{QL_Q}k_{QL_Q}}$, where $i = 1, 2, \cdots, Q$, and $j = 1, 2, \cdots, L_1$
$1 \leq k_{ij} \leq N, k_{11} = 1, k_{QL_Q} = N$

for $i = 1$; $j = 1, 2, \cdots, L_2$ for $i = 2$; $\ldots$; $j = 1, 2, \cdots, L_Q$ for $i = Q$. Design $\{P_1, P_2, \cdots, P_N\}$ as follows: $P_{k_{11}}, \cdots, P_{k_{1L_1}} > \cdots > P_{k_{Q1}}, \cdots, P_{k_{QL_Q}} > 0$, then $V(\rho_f) < V(\rho_{other})$ holds.

**Proof:**

Obviously, $V(\rho_f) < V(\rho_s)$ holds for *N=2*. Assume that for *N-1*, $V(\rho_f) < V(\rho_s)$ is true. Then Eq. (29) holds. For *N*, if $(\tilde{\rho}_f)_{(N-1)(N-1)} < (\tilde{\rho}_f)_{NN}$, design $P_{k_{11}}, \cdots, P_{k_{1L_1}} > \cdots > P_N$, then (31) holds. If $(\tilde{\rho}_f)_{k_{Q1}k_{Q1}} = (\tilde{\rho}_f)_{k_{Q2}k_{Q2}} = \cdots = (\tilde{\rho}_f)_{NN}$, then $NN(\tau) \neq k_{Q1}k_{Q1} \neq \cdots \neq k_{Q(L_Q-1)}k_{Q(L_Q-1)}$ in (30). Design $P_{k_{11}}, \cdots, P_{k_{1L_1}} > \cdots > P_{k_{Q1}}, \cdots, P_{k_{QL_Q}}$, then (31) holds. Proposition 2 is proved. □

Obviously, according to Proposition 1 and Proposition 2, we can obtain Theorem 4.
□

## IV. NUMERICAL SIMULATIONS

In order to verify the effectiveness of the proposed method, consider a three-level system with $H_0$ and $H_1$ as:

$$H_0 = \begin{bmatrix} 0.3 & 0 & 0 \\ 0 & 0.5 & 0 \\ 0 & 0 & 0.9 \end{bmatrix}, H_1 = \begin{bmatrix} 0 & 1 & 0 \\ 1 & 0 & 0 \\ 0 & 0 & 0 \end{bmatrix}, H_2 = \begin{bmatrix} 0 & 0 & 1 \\ 0 & 0 & 0 \\ 1 & 0 & 0 \end{bmatrix} \quad (32)$$

According to $H_0$ and $H_1$, the system is in the degenerate case. Assume that the initial state is an eigenstate as $|\psi_0\rangle = (0 \ 0 \ 1)^T$, and the target state is a superposition state as $|\psi_f\rangle = (\sqrt{2/3} \ -\sqrt{1/3} \ 0)^T$.

According to the design ideas in section II, the control law is designed as $u_k(t) = \gamma_k(t) + v_k(t) + \eta_k, (k=1,2)$. Design $\eta_1 = -0.3771, \eta_2 = 0$ to make the target state $|\psi_f\rangle$ be an eigenstate of $H_0' = H_0 + \sum_{k=1}^{2} H_k \eta_k$. $H_0'$ And design $\gamma_1 = \gamma_2 = \gamma = 0.01 \cdot (\langle \psi | P_\gamma | \psi \rangle - \langle \psi_f | P_\gamma | \psi_f \rangle)$, $v_1(t) = -0.2 \cdot (i\langle \psi |[H_1, P_\gamma]| \psi \rangle)$, $v_2(t) = -0.2 \cdot (i\langle \psi |[H_2, P_\gamma]| \psi \rangle)$, where $P_\gamma = \sum_{j=1}^{3} P_j |\phi_{j,\gamma}\rangle$. According to Theorem 2 in section II, set $P_f = 0.1$ and other two eigenvalues of $P_\gamma$ are 0.4 and 0.6.

In the simulations, the time step size is set as 0.01 a.u., and the control duration is 300 a.u.. The results of numerical simulations are shown in Fig.1 and Fig.2. Fig.1 is the population evolution curves of system, $|c_i|^2, (i=1,2,3)$ is the population of level $|i\rangle$. Fig.2 shows the designed control fields. According to numerical results, we can see that the proposed method is effective.

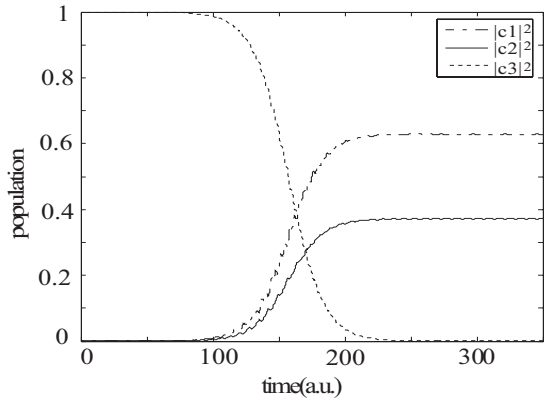
Fig. 1. Population

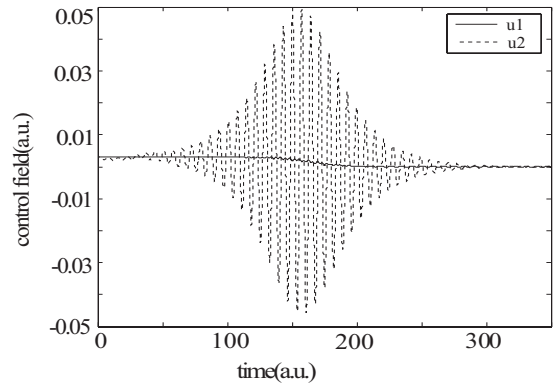
Fig. 2. Control fields

V. Conclusion

In this paper, the Lyapunov control based on the average value of an imaginary mechanical quantity has been improved. By using the proposed method, the quantum Lyapunov control can complete the state transfer task from an arbitrary pure state to an arbitrary pure state for the Schrödinger equation, and from an arbitrary initial state to an arbitrary target state unitarily equivalent to the initial state for the quantum Liouville equation in most cases.